\documentstyle[12pt]{article}
plus2pt

\newcommand{\f}{\begin{equation}}
\newcommand{\ff}{\end{equation}}

\newcommand{\union}{\cup}

\newcommand{\half}{{1\over
2}}

\begin{document}
\thispagestyle{empty}

 \vfil

\centerline{{\Large\bf Gravitons and Loops}}

\vskip.5in

\centerline{ Abhay Ashtekar${}^{1,2}$, Carlo Rovelli${}^{3,4}$ and Lee
Smolin${}^1$}     \vskip
.2in
\centerline{${}^1$ {\it Department of Physics, Syracuse University,
Syracuse
NY 13244-1130, USA}}
\centerline{${}^2$ {\it Max Planck Institut f\"ur Astrophysik, D8046
Garching
bei M\"unchen, Germany}}
\centerline{${}^3${\it Department of Physics, University of
Pittsburgh,
Pittsburgh PA 15260, USA.}}
\centerline{${}^4${\it Dipartimento di Fisica, Universita' di Trento,
Trento;
INFN Sezione di Padova, Italy.}}

\vfil

\centerline{\bf Abstract}
\vskip.1in
\noindent
The recently proposed loop representation is used to quantize
linearized general relativity. The Fock space of graviton
states and its associated algebra of observables
are represented in terms of functionals of
loops. The ``reality conditions'' are realized by an inner product
that is chiral asymmetric, resulting in a chiral asymmetric
ordering for the Hamiltonian, and, in an
asymmetric description of the left and right handed gravitons.
The formalism depends on an arbitrary ``averaging''
function that controls certain divergences, but does not
appear in the final physical quantities. Inspite of these somewhat
unusual features, the loop quntization presented here is completely
equivalent to the standard quantization of linearized gravity.

\vfil
\eject

\section{Introduction}

The standard, perturbative, approaches to quantum gravity begin
with the quantization of linearized general relativity.  They end, of
course, with the discovery that perturbative general relativity is
not renormalizable.  Recently, a nonperturbative approach to
quantum general relativity was introduced
\cite{abhay,tedlee,carlolee,poona,review}.
The framework is based on canonical quantization but has several
new ingredients: the use of (self-dual) connections --rather than
metrics-- as the basic dynamical variables \cite{abhay}, the
introduction of a loop representation of quantum states
\cite{gambini,carlolee}, an extension of the Dirac program of
quantization of constrained systems \cite{poona}, and certain
techniques to deal with diffeomorphism invariant quantum field
theories \cite{tedlee,carlolee}.

In this paper we return to the problem of the quantization of
linearized general relativity and show how the new framework,
which has proved to be useful at the nonperturbative level, may be
applied to this case as well.  More specifically, we present a
quantization of linearized general relativity based on the use of the
new canonical variables \cite{abhay} and the loop representation
\cite{gambini,carlolee}.  This is a necessary step in an ongoing
program whose aim is to resolve the question  of whether the
formulation can lead to a successful quantum
theory of gravity by resolving the short distance problems
nonperturbatively.  More specifically, there
are three reasons that motivate this investigation.

First, the physical interpretation of the mathematical notions
that naturally arise in the exact quantum theory is often rather
obscure. For example, in the full theory, physical quantum states
arise as suitable functions of knot and link classes of closed curves
on a 3-manifold. Compared to the space of states that one normally
encounters in physical theories, this space is rather unusual. In
order to extract the physical content of these states, it would be
extremely useful to relate them to familiar notions such as
gravitons. The first step towards this goal is to recast the Fock
description of graviton also in terms of closed loops. This step is
completed in this paper.

The second motivation is the following. Since all the
techniques used in the non-perturbative approach are novel, it is
important to  test them in situations in which we have good
intuition about the physics of the problem. Over the last two years,
therefore, these techniques have been applied to a number of
examples which mimic various features of full general relativity.
This paper is a continuation of that program. Roughly speaking, the
examples analyzed so far fall into two categories: those that share
certain non-linear features with general relativity, particularly the
presence of constraints which are quadratic in momenta and the
absence of a background space-time metric
\cite{2+1,abhaytate,poona} and those that test the quantization
program in the context of field theories in Minkowski space
\cite{maxwell,yangmills,review}. In this paper, we will be able to
test certain other features of the program which are specific to 3+1
gravity. These include the use of a hybrid pair of canonical variables
in which the configuration variable is complex but the momentum is
real, and the strategy of basing the quantization procedure on the
``loop algebra'' constructed from holonomies of self dual connections
and spatial triads.

The third reason that motivates this work arises from the fact that
two of the postulates of the standard quantization
of free field theories involve notions that cannot be extended to
nonperturbative quantization of diffeomorphism invariant theories:
the use of the positive/negative frequency splitting to define the
vacuum state and operator ordering and the use of the Poincar\'e
group to select the inner product.  In the quantization presented
here, these are replaced by notions that {\it do} extend directly to
non-perturbative general relativity ---the splitting into self dual and
anti-self dual fields rather than positive and negative frequency fields
\cite{aa}, and the use of the ``reality conditions'' rather than
Poincar\'e invariance to select the inner product. The fact that
quantization of the linear theory can be carried out successfully
with these replacements provides further confidence in the program
as a whole.

Our final description of linearized gravity has two curious
features. The first is that the framework has a chiral asymmetry:
the left handed gravitons are described in a different way from
the right handed ones. The origin of this asymmetry lies
in the use of the new canonical variables where the configuration
piece is (the restriction to a spatial slice of) the self dual {\it
or} the anti-self dual part of the spacetime spin-connection. Thus
the asymmetry is not an artifact of the linearization; it is
simply a consequence of replacing the positive/negative frequency
splitting with an ordering prescription that is well defined at both
the linearized and non-perturbative level.  In spite of the
use of one {\it or} the other handed connection, however, the final
theory contains quanta of {\it both} helicities and is completely
equivalent to the standard quantum theory of spin two, rest mass
zero gravitons (essentially because we now allow both the positive
and negative frequency connections \cite{aa}). There exist in the
literature other frameworks aimed at describing the dynamics of
(the real, Lorentzian) general relativity which are also intrinsically
chirally asymmetric: Penrose's twistor program \cite{penrose} and
the Kozameh-Newman formalism based on lightcone cuts
\cite{tedcarlos}. The detailed relation between the three
frameworks, however, remains unclear

The second peculiar feature of our description of linearized gravity
is that in the final picture one has to introduce an arbitrary function
$f_r(\vec x)$ for averaging (or smearing) loop dependent quantities
on a small tube around the loop itself. This is necessary in order to
control the infinities that otherwise arise in certain loop integrals.
(These infinities could not be cured by a standard regularization
technique.) The resulting formalism thus contains a ``momentum
damping''. However, {\it it is exact}: no limit or renormalization are
needed, and the physical quantities do not depend on the specific
choice of the function. Since the technology needed in this averaging
procedure tends to obscure the logic of the rest of the paper, we
will proceed in two steps. First we ignore the averaging and
(formally) construct the entire theory without it. Then we note the
appearance of the divergences and repeat the derivation with the
averaging. This step will simply amount to replacing certain
functions with their averages.

Section 2 discusses the classical features the linearized theory;
the Hamiltonian formulation in terms of the new canonical
variables is recalled and the classical loop algebra is introduced.
Quantization is carried out in section 3; the loop representation is
constructed, the physical states are isolated by solving the quantum
constraints, the correct scalar product is singled out using the
``reality conditions'' and expressions of certain physical
observables are written down explicitly. We conclude in section 4
with a brief discussion of several conceptual issues.

There is a certain number of papers closely related to the present
one. The Hamiltonian formulation of linearized gravity in
terms of the new canonical variables is discussed in detail in
\cite{aajoohan}. The loop representation for Maxwell
electrodynamics is constructed in detail in \cite{maxwell}. This is
the simplest example of a loop representation of a
field theory and therefore well-suited to discuss a number of
subtleties such as the differences between the use of positive
frequency and self dual connections in the construction of the loop
algebra and in the physical interpretation of the basic operators on
loop states. The connection representation of linearized gravity is
constructed and its relation to the loop representation is discussed
in \cite{selfdual}. (These results are reviewed in
\cite{poona,review}.) The truncation of the theory obtained by taking
the limit of Newton's gravitational constant $G$ to  zero, a
truncation related but inequivalent to the linearization  considered
here, is described in reference \cite{googly}.

\section{Classical theory}

This section is divided into four parts. In the first, we recall (from
\cite{aajoohan}) the phase space description of linearized general
relativity in terms of the new canonical variables; in the second we
isolate the true degrees of freedom of the theory; in the third, we
present the reality conditions which need to be imposed to ensure
that we are dealing with the real gravitational field; and, in
the fourth we construct the classical loop algebra for the linearized
theory. This algebra will serve as the point of departure for the
quantum description in the next section.

\subsection{Linearized gravity in new canonical variables}

The new canonical variables for full (real) general relativity
consist of  a pair, $( A_{ai} ,{E}^{ai})$, of fields on a
3-manifold $\Sigma$, where $A_{ai}$ is a complex-valued $SO(3)$
connection and ${E}^{ai}$ is a frame field with density weight
one. (Here $a,b,... = 1,2,3$ are space indices and $i,j,...= 1,2,3$
are internal indices. Unlike in the previous work, for notational
simplicity, we will not use tildes to denote the density weights
since they play a minimal role in this analysis.) To linearize the
theory about flat space,  we choose a background point $({E}^{ai}=
{E}^{ai}_o , A_{ai} = 0)$ in the phase space where
${E}^{ai}_o$ is a flat  triad and consider weak fluctuations
around it. Let us begin with the triad. We set:
\f
{E}^{ai}={E}^{ai}_o +
e^{ai}.
\ff
Since the background triad is flat, so is the metric $q_o^{ab}:=
E^{ai}E^{bj}\delta_{ij}$ constructed from it. For simplicity, we
will use a set of Cartesian coordinates adapted to it so that
we have: ${E}^{ai}_o=\delta^{ai}$. Now the background metric
$q_o^{ab}$ also has components $\delta^{ab}$ and its determinant
is just 1. In what follows, we will often use the background triad to
freely interchange the internal and space indices and suitable
powers of the determinant of the background metric to balance
density weights in various equations. In terms of the perturbed
triad, the usual linear graviton field, $\gamma_{ab}$, that
represents the metric fluctuations around the flat background is
given by: \f
- \gamma^{ab} + \delta^{ab}\gamma^c_{\, \, c} =   2e^{(ab)}
 =: h^{ab}, \label{h}
\ff
where we have defined $h^{ab}$ as the symmetric part of $e^{ab}$.
(The minus sign arises because $e^{ai}$ represents the fluctuation
in the contravariant rather than covariant triad and, the trace
term, because $e^{ai}$ is a density of weight 1.) Let us next
consider the fluctuation in the connection. Since its background
value is chosen to be  zero, we do not need to introduce a new
symbol for its fluctuation. The fundamental (non-vanishing) Poisson
bracket of the linearized theory is
\f
\{  A_{bj}(x),  e^{ai} (y) \}_L = \imath\,  \delta^a_b \delta^i_j
\delta^3 (x,y),
\ff
where, and in what follows, the subscript or the superscript $L$
denotes structures and quantities of the linearized theory.
The linearized theory can now be obtained by expanding the triad
using (1) and keeping in all field equations --the constraints as
well as the equations of motion--  terms which are {\it linear} in
$e^{ai}$ and $A_{ai}$.

Let us begin by recalling the constraints of the full theory. We
have three constraints: a Gauss constraint ${\cal G}^i=
{\cal D}_a E^{ai} \equiv \partial_a {E}^{ai} + G \epsilon^{ijk}
A_{aj} {E}^a_k = 0$, a vector constraint ${\cal V}_a = -i E^{bi}
F_{abi} = 0$, and a scalar constraint ${\cal S} = \epsilon_{ijk}
E^{ai}E^{bj}F_{ab}^k = 0$, where the field strength $F_{ab}^i$ is
defined by: $F_{ab}^i =\partial_a A_b^i -\partial_b A_a^i +
G \epsilon^{ijk}A_{aj} A_{bk}$. The linearization of the Gauss
constraint yields:

\f
{\cal G}^i_L =  (\partial_a e^{ai} + G \epsilon^{ija} A_{aj}) =0.
\label{gauge}
\ff
Note that, in spite of the second term in ${\cal G}^i_L $ the algebra
of these constraints is Abelian:
\f
\{ {\cal G}^i_L (x), {\cal G}^j_L (y) \}_L
=0,
\ff
so that the (internal) gauge group of the linearized theory is
$U(1)\times U(1) \times U(1)$. Note that, by contrast, due to the
fact that ${\cal D}_a = \partial_a  + G A_a \times ... $, in
the full theory we
have:
\f
\{ {\cal G}^i (x), {\cal G}^j (y) \} = G \epsilon^{ijk} {\cal
G}^k (x) \delta^3 (x,y).
\ff
Therefore, the algebra of the linearized theory can be seen as
the  $G  \rightarrow 0 $ contraction of the $SU(2)$ gauge  algebra of
the full theory. To linearize the other two constraints, let us
first define the linearized field strength as:
\f
f_{ab}{}^i := \partial_a A_b^i -
\partial_b
A_a^i;
\ff
Note that it is invariant under the linearized gauge transformation,
generated by the linearized Gauss constraint ${\cal G}^i_L$. The
linearized versions of the vector and the scalar constraints can now
be written as:
\f
{\cal V}_a^L := -\imath f_{ab}{}^b = 0,
\label{dif}
\ff
\f
{\cal S}_L := -\imath  \epsilon^{ab}_{\, \,  \,  \,  c} f_{ab}{}^c
 = 0.
\label{ham}
\ff

The constraint algebra is very simple: all of the constraints
commute with each other. The variations of the fields
$e^{ai}$  and $A_a^i$  under the gauge transformations
generated by the constraints are also simple.
Under the canonical transformation generated by the linearized
Gauss
constraint ${\cal G}(\lambda )_L :=\int d^3x \lambda_i {\cal
G}^{i}_{L} $ we
have
\f
\delta A_a^i = \imath \partial_a \lambda^i, \  \ \  \  \
\delta                 e^{ai} =-\imath
G
\epsilon^{aik}
\lambda_k,
\label{gaugetr}
\ff
whereas under the linearized vector constraint ${\cal V}_L
(\vec{N}) \equiv  \int d^3x N^a {\cal V}^{L}_{a} $ we
have:
\f
\delta A_a^i =0 , \  \ \  \  \  \delta e^a_i = \partial_i N^a
-\delta^a_i        \partial_b N^b .
\ff
Notice that in the last equation it is the {\it internal} index that
becomes the differentiating index in the first term, and that
the second term  expresses the fact that $e^{ai}$ is a
density.  Finally, for the scalar constraint ${\cal S}(N)_L
 \equiv  \int N {\cal S}_L $ we have
\f
\delta A_a^i =0 , \  \ \  \  \  \delta e^{ai} = - 2\epsilon^{bai}
\partial_b N.
\ff

Let us next consider the Hamiltonian. In the full theory, with
asymptotically flat boundary conditions, the Hamiltonian is given
by:
\f
H= \textstyle{1\over 2}\int_\Sigma d^3x N \epsilon^{ijk} F_{abk}
{E}^a_i {E}^{b}_{j} + \textstyle{1\over 2} \oint_{\partial
\Sigma}d^2S_b N
\epsilon^{ijk}A_{ak} {E}^{ai}
{E}^{bj}.
\ff
where $N$ is a density of weight minus one. The linearized
Hamiltonian can be obtained by first choosing the lapse such that the
full evolution keeps the background point $(A_{ai}= 0, E^{ai}=
E^{ai}_o)$ in the phase space  fixed
and then keeping terms which are only quadratic in the  linearized
fluctuations\cite{aajoohan}. The required lapse in the full theory
is just $N= 1/\sqrt{{\rm det} q_o}\equiv 1$.
The linearized Hamiltonian is then given by:
\f
H_L= \int_\Sigma d^3x \ (2\epsilon^{ib}{}_k f_{ab}{}^k e^a{}_i +
G(A_a{}^a A_b{}^b - A_a{}^b A_b{}^a )).
\ff
In this paper we will use the following boundary conditions on the
linearized fields. We assume that $e^{ai}$ falls off as $1/r$, where
$r$ is a radial coordinate in the background metric, while $A_{ai}$
falls off as $1/r^2$. (Usually slightly more stringent conditions
are imposed on the fall-off of $e^{ai}$. (See, e.g.\cite{book}.)
However, our analysis is not sufficiently rigorous to require these
refinements.) These ensure that the integral in (14) converges. Note
however that the expression itself is not invariant under any of the
linearized gauge transformations: the changes it undergoes vanish
only
when the constraints are satisfied. Thus, the Hamiltonian is
unambiguously defined only on the constraint surface. This is of
course a general feature of the Dirac analysis of systems with
first class constraints; the Hamiltonian is unique only up to
addition of constraint functionals. We shall now use this freedom in
the definition to re-express the Hamiltonian in a more
convenient form.

To do so, let us first define the ``magnetic field''  $B^{ci}$ of
the connection $A_a^i$:
\f
{B}^{c i}= {1 \over 2}
\epsilon^{abc}f_{ab}^i.
\label{magnetic}
\ff
Using this definition, going to the Fourier components, and using
the freedom mentioned above, the linearized Hamiltonian can be now
re-expressed in a form that will be  more directly useful later on.
We
have:
\f
H_L = -\textstyle{1\over2} \int d^3k \left (2e_{ba}(-k) - {G \over
k^2}
      B_{ba}(-k) \right) B^{ab}(k) .
\label{hamiltonian}
\ff
Using the reality conditions of section 2.3, one can recognize that
(\ref{hamiltonian}) is just the momentum space version of the
expression $H_L = \textstyle{1\over 2}\int d^3x \bar{A}_a^i(x)
 A^a_i(x)$ in terms of the
self-dual connection and its complex conjugate given in
\cite{abhay}. The fact that the form of the Hamiltonian is
rather curious has nothing to do with gravity; it is a consequence
only of having  written the theory in terms of the self-dual
connection. For example, the Hamiltonian of the Maxwell theory has a
completely analogous form,  $H= \int d^3x \bar{B}^a_+ (B_+)_a $,
when written in terms of the  self-dual ``magnetic field'' $B^a_+ =
 B^a +\imath E^a$ and its complex conjugate \cite{maxwell}.

To complete the classical description, it only remains to specify
the reality conditions. As mentioned above, these will be
discussed in section 2.3.  In the next sub-section, we make a
brief detour to recall  how to extract the physical degrees
of freedom in the Hamiltonian formulation of the classical theory.

\subsection{Physical degrees of freedom}

We review from \cite{aajoohan} the analysis that shows that the
physical degrees of freedom of linearized gravity are
the symmetric, tracefree and transverse parts of $e^{ai}$ and
$A_{ai}$. Although this calculation is not directly used in what
follows, it is nonetheless instructive for two reasons. First, since
the phase space of the linearized theory is the same as that of a
triplet of Maxwell fields, a priori it is not clear that the theory
represents excitations of a single spin-2 field rather than a
triplet of spin-1 fields. It is therefore worthwhile to see how the
presence of the vector and the scalar constraints, which have no
analogs in the theory of spin-1 fields, conspire to extract
precisely a single spin-2 excitation. The second reason is that the
argument given here is essentially reproduced while solving  the
quantum constraint in section 3.3.

To extract the true degrees of freedom, it is simpler (although by
no means essential, see \cite{aajoohan}) to work in the momentum
space. For this, it is convenient to introduce the standard
unit basis vectors $m^a(k), \bar{m}^a(k)$ and
$\hat{k}^a$ satisfying:
\f
m_a(k)k^a =0, \, \, \,  m_a(k)m^a(k) =0 , \,  \,  \,  m_a(k)
\bar{m}^a(k) =    1, \, \, \, \hat{k}^a = {k^a\over
|k|}.
\ff
In terms of these, the metric $q^o_{ab}(k)$ is given by $q^o_{ab}=
\hat{k}_a\hat{k}_b + m_a\bar{m}_b + \bar{m}_am_b$ and the
alternating tensor $\epsilon_{abc}(k)$ will be chosen to satisfy:
$\epsilon_{abc}\hat{k}^am^b\bar{m}^c = -\imath$. Let us now expand
the Fourier transforms of the basic canonical variables in this
basis:

\begin{eqnarray}
A_{ab}(k)&=& A^+m_am_b + A^- \bar{m}_a
\bar{m}_b + A^1 \hat{k}_a m_b +
A^{\bar{1}} \hat{k}_a \bar{m}_b  + A^2 m_a
\hat{k}_b \\
\nonumber  & & +
A^{\hat{2}} \bar{m}_a \hat{k}_b +  A^3 \hat{k}_a \hat{k}_b +
A^4m_a  \bar{m}_b + A^5 \bar{m}_a m_b
\end{eqnarray}
\begin{eqnarray}
e_{ab}(k)&=&e^+m_am_b + e^- \bar{m}_a
\bar{m}_b + e^1 \hat{k}_a m_b +
e^{\bar{1}} \hat{k}_a \bar{m}_b + e^2 m_a
\hat{k}_b + e^{\hat{2}} \bar{m}_a \hat{k}_b   \\
\nonumber
& &  + e^3 \hat{k}_a \hat{k}_b + e^4m_a
\bar{m}_b + e^5 \bar{m}_a m_b.
\end{eqnarray}
The scalar constraint (\ref{ham}) reads $\epsilon^{abi}k_a
A_{bi}=0$, from which we conclude $A^4=A^5$. The three components
of
the vector constraint (\ref{dif}) yield, $A^2 =
A^{\bar{2}}=A^4+A^5=0$. The three components of the gauge
constraint \ref{gauge} yield $e^3=0$, and gives us two
relations
\f
e^1={\imath \over {|k|}}  A^1  , \ \ \   e^{\bar{1}}={\imath \over
{|k|}}  A^
{\bar{1}}.
\ff
Thus, the seven constraints kill five components of the fields and
give us two more relations.

We now impose seven gauge fixing conditions.  As $A_{ai}$
transforms
only under internal gauge transformations, it must be used to gauge
fix ${\cal G}_L^i$. The standard Lorentz gauge condition  $k^a
A_{ai}=0$ then sets $A^3$, $A^1$ and $A^{\bar{1}}$ to  zero, which
then
forces $e^1$ and $e^{\bar{1}}$ to vanish.  Next, let us use the vector
and the
and scalar constraints to fix $e_{ai}$ to be symmetric. However,
note
that only the transverse part of the  antisymmetric piece of
$e_{ai}$ transforms under the canonical transformation generated by
the vector constraint. We have

\f
\delta_{v} e_{[ai]}=\partial_{[i}
v_{a]}.
\ff
Thus, if we impose the condition

\f
\epsilon^{cai}e_{ai}=0,
\ff
only the components of this proportional to $m_c$ and $\bar{m}_c$
fix the linearized vector constraint. To fix the transverse part of
this constraint we impose the trace-free
condition.
\f
e_{aa}=0.
\ff
Finally, the remaining, longitudinal part, which
is
\f
\epsilon^{cai}\partial_c
e_{ai}=0
\ff
gauge fixes the scalar constraint, as one may check
that
\f
\delta_N \epsilon^{cai} \partial_c e_{ai} = - \nabla^2 N .
\ff
These equations set $e^2=e^{\bar{2}}=e^4=e^5=0$, so we are left with
the physical degrees of freedom:
\f
A^{Phy}_{ab} (k) =A^+m_am_b + A^- \bar{m}_a
\bar{m}_b,
\ff
\f
e^{Phy}_{ab} (k) =e^+m_am_b + e^- \bar{m}_a
\bar{m}_b.
\ff
The fields with a $+$ superscript correspond to the positive
helicity and those with a $-$, to negative. Thus, in this
framework, the self dual connection itself carries excitations
of both helicities. This is possible because we have not made a
restriction to positive frequency fields \cite{aa} .

\subsection{Reality conditions}

The new canonical variables for full general relativity are somewhat
unusual in that the configuration variable $A_{ai}$ is complex. To
ensure that one recovers the theory of interest --{\it real} general
relativity-- one has therefore to impose suitable restrictions on
the canonical variables. These are the reality conditions. They
have to be introduced only on the initial data; they are
automatically preserved under time evolution. (For a detailed
discussion of why this is a viable procedure, see \cite{poona} .)
In the full theory, these conditions can be expressed as
\cite{abhay}:
\f
\overline{E^{ai}} =  {E}^{ai},
\ff
\f
A_a^i + \overline{A_a^{i}} = {2\over G} \Gamma_a^i (E),
\ff
where $\Gamma_a^i(E)$ is the spin connection
associated with the frame fields $ {E}^{ck}$. (There exist an
alternate, manifestly polynomial expression of these conditions
\cite{poona}. However, it is not needed for the analysis of the
linearized theory.) If we linearize these expressions, we have, in
terms of the Fourier components,

\f
\overline{e^{ab} (k)} = e^{ab} (-k),
\ff
\f
A_a^i(k) + \overline{{A}_a^{i}(-k)} = {2\over G} \imath
\epsilon^{bci}(k_c e_{(ab)}(k) -\textstyle{1\over 2} k_c
\delta_{ab}e(k) -\textstyle {1\over 2} k_ae_{bc}(k)).
\ff
In the quantum theory we will only need to impose the reality
conditions on the physical degrees of freedom. In the notation of
the last section, these conditions are:

\f
\overline{{e}^\pm (k) }= e^\pm (-k) ,
\label{rr1}
\ff
\f
A^\pm(k) + \overline{{A}^\pm (-k)} = - \pm {2\over G} |k| e^\pm (k).
\ff
For later use, let us recast this last expression in terms of the
magnetic fields. For the physical components we find

\f
B^\pm (k)= \pm |k| A^\pm (k),
\label{sign}
\ff
which gives
us,
\f
B^\pm (k)+ \overline{{B}^\pm (-k)}= -2 { k^2 \over G} e^\pm(k)  .
\label{rr2}
\ff
This completes the summary of the phase space formulation of
the linearized theory in terms of new canonical variables.

\subsection{Linearized classical loop algebra}

The loop representation in full quantum general relativity was
constructed as a linear representation on the vector space of
(regular) functionals on the loop space of a certain algebra of
(internal) gauge invariant operators, known as the loop algebra
\cite{carlolee}.  In this paper, we want to introduce the analogous
construction for the linearized theory. In this subsection, we will
complete the first step by defining the appropriate classical
loop variables and presenting their Poisson algebra.

The first important property of the loop variables is that they are
invariant under the internal gauge transformations. The simplest
loop variables which are invariant under the linearized ($U(1)\times
U(1)\times U(1)$) transformations generated by ${\cal G}_L$ are
\f
t^i[\gamma ] \equiv e^{ G \oint_\gamma A_a^i \dot{\gamma}^a ds }.
\label{t}
\ff
This is an Abelian holonomy, since the connection is now Abelian.
Note that the internal index has become a label on the loop
variable $t^i[\gamma]$.  Since the connection is left
untouched by the canonical transformations generated by
the vector and the scalar constraints, it follows immediately
that these loop variables commute also with all the remaining
constraints and are thus physical observables of the
linearized theory.

The $t^i[\gamma]$ satisfy several relations, by virtue of their
definitions. First, they are invariant under monotonic
reparametrizations of the loops. Second, if $\alpha $ and
$\beta$ are two loops that meet at a common base point, we
have
\f
t^i[\alpha ] t^i [\beta ] = t^i [\alpha \circ \beta ],
\label{abelianident}
\ff
where $\alpha \circ \beta $ is the loop obtained by combining
$\alpha$ and $\beta $ (using a line segment to join them if they do
not intersect). Finally, if $l $ is an open line, we
have the ``retracing'' identity
\f
t^i[\alpha \circ l \circ l^{-1} ] = t^i [\alpha ].
\label{retracing}
\ff

To express the dynamics of the linearized theory in terms of the
loop variables, we need to write the ``magnetic field'' (15) as a
function of the holonomy $t^i[\gamma ]$. This can be done using a
limiting procedure. For each point $x$ , a unit vector $\hat{e}^c$
at $x$ and a real number $\delta$, we define a loop
$\gamma^c_{x,\delta}$, which is a circle of radius $\delta$ in the
plane perpendicular to $\hat{e}^c$ and centered at $x$. Then, as
in Maxwell theory, the ``magnetic field'' $B^{ci}(x)$ is given by :
\f
G \ {B}^{ci}(x) = {\rm lim} {1\over \pi\delta^2}
(t^i[\gamma^c_{x,\delta}] -1).
\label{mag}
\ff

This concludes the discussion of the classical configuration
loop variables. Let us now introduce the momentum loop variables.
In the full theory, these momenta are ``handed'' loop variables,
associated with oriented loops. In the linearized theory, the
momentum
variables are considerably simpler. From the form of the internal
gauge transformation (\ref{gaugetr}) generated by the linearized
Gauss constraint, it is clear that $h^{ab}:= 2e^{(ab)}$ is gauge
invariant. Therefore, we can just take it to be our momentum
variable. Note that the origin of this simplicity is the Abelian
nature of the present internal gauge group. In the full theory,
by contrast, the group is non-Abelian and the construction of the
momentum loop variables is more involved. Note that, since they
depend only on the triads, these momenta have vanishing Poisson
brackets among themselves. Thus, the only non- zero bracket
between
the $t^i[\gamma]$ and $h^{ab}(x)$ are:
\f
\{h^{ab} (x), t^i [\gamma ] \}_L =  -2\imath G\  (\oint ds\ \delta^3
(x,\gamma (s))\
\dot{\gamma}^{(a}  \delta^{b)i}) \ \  t^i [\gamma ] \ ,
\label{alg}
\ff
where no sum is performed over $i$ in the right hand side. The
variables $h^{ab}$ and $t^k [\gamma ]$ will be called the {\it
linearized loop variables} even though now the momentum variable is
not associated with a loop. The vector space generated  by them is
clearly closed under the Poisson bracket. This Poisson algebra will
be called the {\it linearized loop algebra}. In spite of this
apparent difference between the construction of loop variables in
the full and the linearized theory, it does turn out that the
linearized loop algebra can be recovered from the loop algebra of
the full theory through an appropriate limiting procedure
\cite{linearization}.

For later use, let us decompose $h^{ab}$ in Fourier components
\f
h^{ab}(k) = {1\over (2\pi)^{3\over 2}} \int d^3 x\ e^{\imath k \cdot
x }\ h^{ab} (x).
\ff
The basic Poisson bracket of the linearized theory can now be
written as:
\f
\{h^{ab} (k) , t^i [\gamma ] \}_L = -2\imath G\  (F^{(a} [k,\gamma ]
\delta^{b)i})\ \  t^i [\gamma ]
\label{algebra}
\ff
where $F^a[k, \gamma ]$ is given by:
\f
F^a [k,\gamma ] = {1\over (2\pi)^{3\over 2}} \oint ds\ e^{\imath k
\cdot \gamma (s)}\                   \dot{\gamma}^{a}(s).
\label{ff}
\ff
This quantity plays a crucial role in the description of the theory
that follows; it will be called the {\it form factor} of the loop
$\gamma$.  The form factor is automatically transverse:
\f
k_a F^a [k,\gamma ] =0,
\label{transv}
\ff
and it can be thought of as the Fourier transform of a distribution
having  support on the loop itself:
\f
 F^a [k,\gamma ] = {1\over (2\pi)^{3\over 2}} \int d^3x\
e^{ikx}\ F^a[x,\gamma],
 \ff
\f
F^a [x,\gamma ] =  \oint ds \ \dot\gamma^a(s) \delta^3(x,\gamma(s)).
\label{loopdistr}
\ff
Finally, $F^a[k,\gamma ]$ have a dual interpretation. First,
if one fixes a loop $\gamma_o$ then $F^a[k, \gamma_o ]$, regarded
as a function of $k$, has the full knowledge of the spatial structure
of (the holonomic equivalence class of) the loop $\gamma_o$.
This is  why, following the terminology used in particle physics, we
have called $F^a[k, \gamma]$ a form factor. However, if one fixes
a momentum vector $k_o$, then $F^a [k_o ,\gamma ]$, regarded as
functions of $\gamma$, are just the logarithms of holonomies
of certain connections. To see this, note first that since $F^a[k,
\gamma ]$ is transverse, it has only two components:
\f
F^a [k,\gamma ] =  F^+ [k, \gamma ] m^a(k) + F^- [k, \gamma]
\bar{m}^a (k).
\ff
$F^{\pm}[k_o,\gamma ]$ is the logarithm of the holonomy of the
plane
wave  connection with wave vector $k_o$ and positive (respectively,
negative) helicity. This dual interpretation makes several
mathematical properties of form factors transparent. We will see in
the next section that these form factors  play an important role in
quantum theory.

\section{Quantum Theory}

This section is divided into seven parts. In the first, we introduce
the loop representation for the linearized theory. In the second,
we make a mathematical digression to show how the form factors
introduced in section 2.4 can be used to select the appropriate class
of ``regular'' functions on the loop space. In the third, we present
the solutions to the quantum constraints and in the fourth we show
how the inner product can be selected on the space of physical
states using reality conditions. The fifth subsection gives the
expression of the quantum Hamiltonian operator in the loop
representation and provides the physical interpretation to various
operators and states. A part of the construction up to this point is
formal in that it contains certain divergent integrals. These are
discused in the sixth subsection. The seventh shows how these
divergences can be avoided by introducing suitable smearing
procedures. As explained in the Introduction, we could have
introduced the smearing right in the beginning of this section
thereby avoiding the issue of divergences altogether. We chose
not to do so only in order to simplify the presentation.

\subsection{Linearized loop representation}
The main idea of the loop representation is to construct the quantum
loop algebra starting from the classical one and then find its
representation using, for the carrier space, a space of functions
over a loop space. (Note that in such a representation the quantum
states are {\it not} functions on a
configuration space. A general discussion on the loop
representation can be found in \cite {carlolee,poona,review}).
The construction of the quantum loop algebra is straightforward.
It is generated by the quantum holonomy operators
$\hat{t}^i[\gamma ]$ and their ``momenta''
$\hat{h}^{ab}(k)$ which, for brevity, will be called the ``graviton
operators''. The basic (non vanishing) commutation relation is:
\f
[\hat{h}^{ab}(k) ,  \hat{t}^i[\gamma ] ] = 2G\hbar
(F^{(a}[k,\gamma ]\ \delta^{b)i}) \ \ \hat{t}^i[\gamma ],
\ff
where, as before, $F^a[k,\gamma ]$ is the form factor of the loop
$\gamma$. A new issue does arise, however, in the construction of
the representation of this algebra. For, unlike in the loop
algebra of the full theory, the holonomy operators
$\hat{t}^i[\gamma ]$ now carry an internal index. Consequently, we
can no longer just repeat the procedure used in \cite{carlolee}.
To address this issue, a new strategy is needed. In this paper, the
solution we propose involves a change in the carrier space of the
representation itself: in place of functions of loops, we now use
functions on the space of triplets of loops. Roughly, the reason
for this choice is the following. Kinematically we can think of
linearized gravity in the new variables as three copies of
Maxwell theory, labelled by the index $i$. The appropriate
carrier space for the Maxwell loop algebra is a space of functions
of (single) loops \cite{maxwell}. Since, in general, the appropriate
space of quantum states for a composite system is the tensor
product of the space of states for the single system, it is now
natural to replace the loop space with the triple Cartesian product
of this space with itself.

Let us begin with some definitions. By a {\it single loop} we mean
a piecewise smooth closed curve $\alpha$ in $R^3$.  A
{\it multiple loop} $\eta=\{\alpha_1 \cup \alpha_2 \cup  ...
\cup \alpha_n\}$, will stand for a collection of a finite number of
single loops.  We use the same notation (a Greek letter) for multiple
loops and for single loops (single loops being considered a
special case of multiple ones). The line integral around a
multiple loop $\eta$  is defined to be the sum of the line
integrals around its components
$\alpha_n$:
\f
 \oint_\eta f_a\ \dot\eta^a ds \equiv \sum_n  \int_{\alpha_n}
f_a\
(\dot\alpha_n)^a ds .
\ff
Next, let us identify two single loops $\alpha$ and $\beta$ if they
have the same support and the same orientation; or, if they are
related by a retracing identity $\beta = \alpha \circ l \circ
l^{-1}$ (see equation (\ref{retracing})). In addition, let us
identify two multiple loops $\eta$ and $\rho$ whenever  (see
equation (\ref {abelianident})

\f
       \eta = \{ \alpha \cup \beta \cup \gamma \cup ...\delta \}, \ \ \ \
  \rho= \{     \alpha \circ \beta \cup \gamma \cup ...\delta \}.
\ff
The space of (oriented) multiloops with these identifications
will be called the $U(1)-${\it holonomic loop-space} (or, simply
loop space if there is no ambiguity), and denoted ${\cal
HL}$. (This space has also ben called the non-parametric
loop space by Gambini and Trias \cite{gambini}.)  The reason for
the name ``holonomic loop space'' is that the identifications
realize the relations induced on loops by the properties
of $U(1)$ holonomies. An important consequence of these
identifications is that {\it any multiloop is equivalent
to a single loop}. Thus, strictly speaking, it would have sufficed
to work with single loops; our present description has a redundancy.
However, as we will see below, many of the basic formulas are
easier to express in terms of multi-loops. As emphasized and
exploited by Gambini and Trias \cite{gambini}, ${\cal HL}$
has a natural group structure.

Let us next consider the space ${\cal HL}^3 = {\cal HL} \times {\cal
HL} \times {\cal     HL}$ formed by triples of multiple loops. An
element of ${\cal HL}^3$ is denoted
$\vec\eta$:
\begin{eqnarray}
\vec{\eta} & = & \{\eta^1 , \eta^2 , \eta^3 \} \\
\nonumber
 & = &   \{ \ \{\alpha^1_1 \cup \alpha^1_2\cup  ... \cup
\alpha^1_{n_1}\} , \{
\alpha^2_1\cup \alpha^2_2\cup ... \cup \alpha^2_{n_2}  \} , \{
\alpha^3_1\cup \alpha^3_2\cup  ... \cup \alpha^3_{n_3} \} \ \} .
\end{eqnarray}
The $i$-th multiple loop in the triple is denoted $\eta^i$. For
simplicity, we will use the same term, {\it loop}, to denote
single loops, multiple loops, or triplets of multiple loops; the
context will make the precise meaning clear. Note also that ${\cal
HL}^3$ can also be seen as the space of the collections of single
loops of three kind (three colors): $\eta^1, \eta^2 $ and
$\eta^3$. The space ${\cal HL}^3$ is naturally equipped with an
operation, the union:
\f
\vec{\eta} \union \vec{\rho}
 :=
 \{ \eta^1 \union
\rho^1 , \eta^2 \union \rho^2 ,
\eta^3 \union \rho^3
\}.
\ff
It is convenient also to define union of a specific kind,
$\union_i$, of an element of ${\cal HL}^3$ with a single loop. For
example, we have: \f
\vec{\eta} \union_1 \alpha := \{ \eta^1
\union \alpha ,\ \eta^2 ,\  \eta^3 \}.
\ff

With these preliminaries out of the way, we are now ready to
introduce the loop representation. Th carrier space will be the
space ${\cal S}$ of (suitably regular) functionals on ${\cal HL}^3$.
Thus, each $\psi\in {\cal S}$ has the
form:
\f
\psi [\vec{\eta}]  =  \psi[\eta^1 , \eta^2 , \eta^3 ] \ . \ff
The regularity conditions will be specified in the next subsection.
The representation map is defined as follows. The action of the
holonomy operator $\hat{t}^i[\gamma ]$ is given by :
\f
\left ( \hat{t}^i[\alpha] \psi \right )
[\vec{\eta}] :=  \psi^\prime [\eta ]=  \psi[
\vec{\eta} \cup_i {\alpha}].
\label{holoop}
\ff
Thus, the operator $\hat{t}^i[\alpha]$, operating on a state
$\psi$ produces a new state $\psi^\prime$ whose value,
on the loop $\vec{\eta}$, is the value of the old state $\psi$ on
the loop $\vec{\eta} \cup_i {\alpha}$. (This operator is analogous
to or translation operator in wave mechanics: $(U(a)\psi)(x) :=
e^{\imath a \hat p}\psi(x)=\psi(x+a)$). The second basic loop
operator is the graviton operator $\hat{h}_{ab}(k)$. Its action
will be given by:
\f
(\hat{h}^{ab}(k) \psi) [\vec{\eta}]=2 \hbar G\
F^{(a}[k,\eta^{b)}]\
 \psi
[\vec{\eta}]
\label{gravitonop}
\ff
where $\eta^b$ is the $b$'th element of the triple of loops
$\vec{\eta}$. Thus, $\hat{h}^{ab}(k)$ is in fact diagonal in the loop
basis (and therefore analogous to the $\hat{x}$ operator of wave
mechanics). A straightforward calculation shows that the
operators $\hat{t}^i[\gamma ]$ and $\hat{h}^{ab}$ so defined satisfy
the required commutation relations; we do indeed have a
representation of the quantum loop algebra. This completes the
kinematics of the quantum theory.

To define the dynamics --i.e. to specify the constraint and the
Hamiltonian operators-- we need the operator corresponding to the
the linearized magnetic field (\ref{magnetic}).  By using its
expression in terms of the loop observables, equation (\ref{mag}),
we
have
\begin{eqnarray}
\hat{B}^{ci}(x) \psi [\vec{\alpha}] &
\equiv
&  {1 \over G}
\lim_{\delta  \rightarrow 0} {1 \over \pi\delta^2}
\left (\hat{t}^i[\gamma^c_{x,\delta} ]
-1  \right )  \psi [\vec{\alpha }]  \\
\nonumber
& = & {1 \over G} \lim_{\delta \rightarrow 0}
{ 1\over  \pi\delta^2} \left (
\psi[ \vec{\alpha} \cup (\gamma^c_{x,\delta})^i ]
-\psi [\vec{\alpha} ]   \right )    \\  \nonumber
&\equiv  &  {1 \over G}\  {\delta\over\delta\gamma^{ci}(x)}\
\psi [\vec{\alpha} ]
{}.
\label{der}
\end{eqnarray}
In the last step we have introduced the notation ${\delta /\delta
\gamma^{ci}(x)}$  to emphasize that the operator $\hat{B}^{ai}(x) $
can be regarded as a derivative operator in the holonomic loop-
space.
Note, however, that this operator is {\it not} a functional
derivative. Rather, it is closely related to the operator
$\Delta_{\mu\nu}(P)$ introduced by Gambini and
Trias.\cite{gambini}
Thus, our treatment of the magnetic field operator differs from
treatment one generally finds in gauge theories on loop spaces
where
the field operators {\it are}  represented by derivative operators.
This change of strategy is crucial for the case of full gravity,
where the action of the limit that defines the derivative may not be
well defined on diffeomorphism invariant state \cite{carlolee}.

The linearized scalar and vector constraints can be expressed
in terms of the magnetic field operator. Let us first
take the Fourier transform.
\f
G\  \hat{B}^{ci}(k)\psi
[\vec{\alpha}] = {\delta \over
{\delta \gamma^{ci}(k)}}   \psi [\vec{\alpha}] \equiv {1\over
(2\pi)^{3\over 2}}\int d^3x e^{\imath k \cdot x } {\delta \over
{\delta \gamma^{ci}(x)}}   \psi [\vec{\alpha}].
\label{mfoft}
\ff
The linearized constraint operators can then be expressed
as:
\f
\hat{\cal V}^L_a (k) = {i\over G} \epsilon_{aci}
{\delta \over \delta \gamma^{ci}(k)}
\label{qdif}
\ff
and
\f
\hat{\cal S}^L(k) = -{2i\over G}{\delta \over \delta
\gamma^{\, c}_{c}(k)}.
\label{qham}
\ff
We will see later that on the space of regular loop states defined
in the next subsection, the action of the magnetic field operator,
and hence also of the constraint and the Hamiltonian operators,
can be expressed in a simple way in terms of loop form factors
introduced in section 2.4.

\subsection{Regularity conditions on loop states}

We now complete the introduction of the linearized loop
representation by specifying the regularity conditions that need
to be imposed on the loop functionals $\psi(\vec\alpha)$.

For this purpose, we will use the loop form factors $F^a[k ,\gamma
]$. Note first that $F^a[k,\gamma ]$ are well-defined on the
$U(1)-$holonomic loop space $\cal HL$ because two equivalent loops
have the same form factors. (Recall that the form factors
themselves
can be regarded as $U(1)$ holonomies of certain connections.) Since
the Fourier transform of the form factor is a distribution with
support on the holonomic-loop itself, it is clear that, as a
function of $k$, the form factor of a given holonomic loop $\eta$
uniquely determines $\eta$. Thus, seen as functions of $k$, the form
factors serve as good labels for holonomic-loops. They cannot
be regarded as coordinates in the technical sense because the space
of form factors itself does not form a linear space; if
$F^a[k,\gamma^i ]$ is the form factor of $\gamma^i$, in general
$\lambda F^a[k, \gamma^i]$ will not be a form factor of any loop.
Nonetheless, we have the following interesting mathematical
structure.
Consider the space ${\cal F}^3 :={\cal F}\times {\cal F}\times {\cal
F}$,  where ${\cal F}$ is the space of vector fields
$F^a(k)$ in the momentum space satisfying $\overline{F^a(k)} =
F^a(-k)$ and $F^a(k)k_a = 0$. The first condition ensures that
$F^a(k)$ is the Fourier transform of a {\it real} vector field
$F^a(x)$ while the second ensures that $F^a(x)$ is divergence free. (A
more precise definition of this function space will be given later.)
Form factors $F^a[k ,\gamma^i]$ can be now regarded as
1-1 mappings from the loop space ${\cal HL}^3$ {\it into} the space
${\cal F}^3$. Therefore, any functional $\phi (F^{ai}(k))$ on ${\cal
F}\times {\cal F}\times {\cal F}$ determines uniquely a functional
on loop space via
\f
\psi[\vec\gamma] =
\phi[F^{ai}(k)]|_{F^{ai}(k) = F^a(k, \gamma^i )}
\label{f}
\ff
and {\it any} loop functional can be obtained in this way. Thus, the
space of the functionals $\phi$ on ${\cal F}^3$ is mapped linearly
onto the  space of the loop functionals. We now exploit this fact to
specify the regularity conditions on the loop functionals $\psi
(\vec\gamma )$ in terms of those on the functionals $\phi (F^{ai}
(k))$.

A loop functional $\psi (\gamma)$ will be said to be {\it regular}
--and thus belong to the carrier space of the representation-- if
and only if it is obtained via ({\ref{f}}) from a
functional $\phi$ which admits a convergent Taylor
expansion on the space ${\cal F}^3$. Thus, regular loop states arise
from analytic functionals on ${\cal F}^3$. Let us give a few
examples.
A linear functional on ${\cal F}^3$ is of the type $\phi(F^{ai}(k)) =
\int d^3k\ C_{ai}(k) F^{ai}(k)$, while a $n$-th order functional is
of the type $\int d^3k_1 ....\int d^3k_n \    C_{a_1i_1 ...a_ni_n}
(k_1,...k_n)
F^{a_1i_1}(k_1) ....F^{a_n i_n}(k_n)$. Any regular loop
state $\psi(\gamma)$ is obtained via ({\ref {f}}) from a (possibly
infinite, convergent) linear combination of these n-nomials on
${\cal F}^3$. Note that the linear functionals have six degrees of
freedom for each momentum  value $k$, since it is only the
transverse part of $C_{ai}(k)$ in the index $a$ that contributes to
the integral. We will see that the imposition of the (four) scalar
and vector constraints reduce the  degrees of freedom  precisely
to two. These will correspond to the two degrees of freedom of the
graviton discussed in section 2.2. (A key feature of the loop
representation is that the quantum Gauss constraint does not have
to be imposed; it is automatically
satisfied. This is a very general  property. It holds also in full
general relativity, as well as in the Maxwell and Yang-Mills
theories.)

Note that since the map from ${\cal HL}^3$ to ${\cal F}^3$ is into
and not onto (i.e. injective and not surjective), the linear mapping
({\ref{f}}) has a non-trivial kernel. In various calculations,
we will need to express the action of the loop operators on states
$\psi (\vec\gamma )$ using form factors $F^{ai}(k)$. Therefore, it is
important to ask if the restriction of the map ({\ref{f}}) to
polynomials $\phi (F^{ai}(k))$ is faithful: If it is not, we
will not be easily able to go back and forth between operators on
loop states $\psi (\vec\gamma )$ and those on polynomial
functionals
$\phi(F^{ai}(k))$ on ${\cal F}^3$.  Fortunately, it turns
out that the kernel of the restricted map is in fact trivial. We can
therefore pass freely between the regular loop states $\psi
(\vec\gamma)$ and the polynomial functionals $\phi (F^{ai}(k))$ on
${\cal F}^3$ from which they arise.

To prove this assertion, we proceed in two steps. First we show
that distinct n-nomials of the same degree give rise
to distinct loop functionals. Let us begin with degree one, i.e.,
linear functionals. The problem then reduces to showing that, if
for every loop $\alpha$ in ${\cal HL}$ we have
\f
\int d^3k \ \ c_a(k) F^a[k,\alpha] = 0,
\label{c}
\ff
then for {\it every} $F^a(k)$ in ${\cal F}$, we must have:
\f
\int d^3k \ \ c_a(k) F^a(k) = 0.
\label{cc}
\ff
This is easy to establish. Equation (\ref{c}) implies:
\f
 \int d^3k c_a(k) \oint ds e^{\imath k \alpha(s)}
\dot\alpha^a(s) =  \oint ds \dot\alpha^a(s) c_a(\alpha(s)) =
\oint_\alpha dS^a c_a = 0,
\ff
where $c_a(x)$ is the Fourier transform of $c_a(k)$.  Since its line
integral on any loop is  zero, $c_a(x)$ is curl free, whence $c_a(k) =
k_a c(k)$. Inserting this in (\ref{cc}), we obtain the required
result because every $F^a(k)$ in ${\cal F}$ is transverse. Let us
next consider a quadratic functional on ${\cal F}$. Let us suppose
that the restriction of this functional to the $F^a(k)$ that arise
from form factors of arbitrary loops vanishes. Then, we have, for
all  single loops $\alpha$ and $\beta$:
\f
\int d^3k_1 \int d^3k_2 c_{ab}(k_1,k_2) F^a[k_1,\alpha]
F^b[k_2,\alpha] = 0 ,
\label{ccc}
\ff
and, considering a multiple loop formed by these two single
loops, we also have
\f
 \int d^3k_1 \int d^3k_2 c_{ab}(k_1,k_2)
F^a[k_1,\alpha\cup\beta] F^b[k_2,\alpha\cup\beta] =
0.
\ff
Using the fact that the form factor of the union of two loops
is the sum of the form factors, we conclude, from the last two
equations,
\f
 \int d^3k_1 \int d^3k_2 c_{ab}(k_1,k_2) F^a[k_1,\alpha]
F^b[k_2,\beta] = 0.
\ff
Now we can repeat the argument used in the linear case, and
conclude  that $c_{ab}(k_1,k_2) = (K_1)_a (k_2)_b c(k_1,k_2)$,
whence the quadratic functionals have to vanish everywhere on
${\cal
F}$. It is clear that the discussion can be repeated for an
arbitrary n-nomial: If the restriction of an n-nomial on ${\cal F}$
to the image of the loop space ${\cal HL}$ vanishes, then the
n-nomial must vanish everywhere on ${\cal F}$. Finally, it is
straightforward to replace in this result ${\cal F}$ by ${\cal F}^3$
and ${\cal HL}$ by ${\cal HL}^3$.

The second step in the argument uses this fact to establish that
the same result holds for an arbitrary linear combination of the
n-nomials, i.e., an arbitrary polynomial. That is, we have to show
that the restrictions to the image of ${\cal HL}$ of functions
which are n-nomials of {\it different} oder are necessarily linearly
independent. To carry out this step, we introduce the following
linear operator on the space of loop states:
\f
\hat P \psi[\vec\alpha] :=
\psi[\vec\alpha\cup\vec\alpha].
\ff
It is straightforward to check that every n-nomial functional $\phi
(F^{ai}(k))$ of degree $n$ has the property that if $\psi
(\alpha^i) = \phi (F^a(k,\alpha^i))$, then:
\f
\hat P\ \psi (\vec\alpha ) = 2^n  \psi(\vec\alpha).
\ff
Consequently, the restrictions to the image of ${\cal HL}^3$ of
n-nomial
with different degree belong to the eigenspaces of $\hat{P}$ with
different eigenvalues. Therefore they cannot be linearly dependent.
(To see this, note that, since the eigenvalues are all real,
one can introduce an inner-product with respect to which the
operator
$\hat{P}$ is Hermitian and hence these distinct eigenspaces are
mutually orthogonal. In fact, the operator is
closely related to the number operator in quantum theory.)

Thus, we have shown that the space of regular functionals on the
loop space is indeed isomorphic with the space of polynomials
on a linear space, ${\cal F}^3$. This fact will be used repeatedly
in the calculations that follow. Finally, note that we have
purposely left the precise definition of the function space ${\cal
F}$ open at this stage. The appropriate function space will be
singled out by physical considerations involving the inner product
and the Hamiltonian in section 3.7.

\subsection{Solutions to the constraints: physical states}

Let us now impose the linearized scalar and vector constraints on
the
(regular) loop states. That is, let us find the subspace ${\cal
S}_{Phy} \subset {\cal S}$ annihilated by the operators $\hat{\cal
V}^L_a$ and $\hat{\cal S}^L$ defined in (\ref{qdif}) and
(\ref{qham}).

To carry out this task, it is convenient to first evaluate the
action of the magnetic field operator $\hat{B}^{ai}(k)$ on regular
loop states. If $\psi (\vec\alpha ) = \phi (F^a (k, \alpha^i))$,
then using the definition (\ref{der}) of the operator $\delta /\delta
\gamma^{ai}(k)$, the expression (\ref{mfoft}) of the  magnetic field
operator simplifies to:
\f
G\ \hat{B}^{ci}(k) \cdot \psi [\vec\alpha ] =
\int d^3p  {\delta \phi \over {\delta F^{bj}(k)}}\ \
{\delta F^b(p ,\gamma^j)\over {\delta \gamma^{ai}(k)}}\  .
\ff
Now, from the definition (\ref{ff}) of form factors, it is
straightforward to evaluate the second term. We have:
\f
{\delta \over { \delta \gamma^{ai}(k) }} F^b [p, \gamma^j ]
=  \imath\  \delta^3 (k+p) \ \epsilon^{bac}\  k_c\  \delta^{ij},
\ff
Substituting this result in the above expression of the magnetic
field operator, we obtain:
\f
G\hat{B}^{ci}(k)\cdot \psi[\vec\alpha ] = i C^{ai}(-k)
\epsilon_a{}^{cd} k_d \ \ ,
\label{mfo}
\ff
where we have set
\f
 C_{ai}(k) :={\delta\phi \over\delta F^{ai}(k)}.
\ff
Equation (\ref{mfo}) now implies the following necessary and
sufficient conditions for $\psi [\vec\alpha]$ to be a physical
state:
\f
\hat{\cal S}^L \psi [\vec{\alpha}] =0  \  \ {\rm iff}
\  \   \epsilon^{aic} k_c C_{ai}(k) = 0
\ff
and
\f
\hat{\cal V}^L_e \psi [\vec{\alpha }] =0 \  \ {\rm iff}
\   \  k^c C_{ec}(k) - k_e C_a^{\, a}(k) = 0.
\ff
Note, furthermore, that, because $F^{ai} (k)$ in ${\cal F}^3$ are
all transverse, the functional derivative of $\phi$ with respect to
them is determined only up to an additive term of the type
$k_av_i(k)$. Using this freedom, without any loss of generality, we
can restrict ourselves to those $C_{ai}(k)$ which
satisfy:
\f
k^a
C_{ai}(k)
=0.
\ff
We now have seven equations for the nine components of
the $C_{ai}(k)$. The solutions have two degrees of freedom and, not
surprisingly, are symmetric, tracefree transverse fields:
\f
C_{ab}(k) = A^+m_a(k) m_b(k) + A^- \bar{m}_a(k) \bar{m}_b(k).
\ff
It follows that the  general solutions are those that are
functions only of the tracefree transverse components
of the $F^a[k,\alpha^i]$, which are
\f
F^+ [k,\vec{\alpha}] \equiv \bar{m}_a
\bar{m}_{i}
F^a[k,\alpha^i],
\ff
\f
F^-[k,\vec{\alpha}] \equiv m_a
m_i
F^a[k,\alpha^i].
\ff
That is, the general solution to the quantum constraints has the
form
\f
\psi[\vec\alpha]=\phi[\,
F^+[k,\vec\alpha],
F^-[k,\vec{\alpha}]\, ].
\label{solution}
\ff
The space spanned by these functions will be denoted
${\cal S}_{Phy}$.

Finally, one can repeat the arguments presented in section 3.2 to
show that there is a 1-1 correspondence between physical states
$\psi [\vec\alpha ]$ and (analytic) functionals $\phi$ on  ${\cal
F}^3_{Phy} := {\cal F}^+ \oplus {\cal F}^- $, where ${\cal F}^\pm$ are
spanned by functions $F^\pm(k)$ given by: $F^+(k) = F^{ai}(k)
\bar{m}_a(k)\bar{m}_i(k)$ and $F^-(k)= F^{ai}(k){m}_a(k){m}_i(k)$.
The spaces ${\cal F}^\pm$ can be defined in their own right (without
reference to ${\cal F}^3$) as the vector spaces of functions
$F^\pm(k)$ on the momentum space with spin weight $\pm 2$ which
satisfy the reality conditions $\overline{F^\pm (k)} = F^\pm (-k)$.
Thus, just as we had, in section 3.2, a representation of the
regular loop states by functionals on a linear space ${\cal F}^3$,
we now have a representation of the physical loop states by
functionals on a linear space ${\cal F}^3_{Phy}$. We will use this
representation to simplify the expressions of the physical
operators. The physical operators --i.e. loop operators which leave
the physical space ${\cal S}_{Phy}$ invariant-- are easy to
identify: they are $\hat{h}^\pm (k)$ and $\hat{B}^\pm(k)$, obtained
by contracting $\hat{h}^{ab}(k)$ and $\hat{B}^{ai}(k)$ with the
vectors $m^c(k)$ and $\bar{m}^c(k)$ in the obvious way. Their action
on physical states can now be expressed as follows. For all $\psi
[\vec\alpha ]$ in ${\cal S}_{Phy}$ with $\psi [\vec\alpha ] = \phi
[F^\pm (k, \vec\alpha )]$ , \f
\hat{h}^\pm (k)\cdot\psi [\vec \alpha ] = 2 \hbar G F^\pm
(k,\vec\alpha)\ \ \psi[\vec\alpha]
\label{hop}
\ff
\f
\hat{B}^\pm(k) \cdot \psi [\vec\alpha ] = \pm {|k|\over G}
[{\delta \over {\delta F^\pm (-k)}}\cdot \phi (F^\pm (k))]|_{ F^\pm
(k) = F^\pm (k ,\vec\alpha )}\ \ .
\label{bop}
\ff
\subsection{Physical inner product}

Up to this point, the space of states has the structure of only a
complex vector space; it is not equipped with an inner product.
The idea of postponing the introduction of the inner product until
after one has solved the quantum constraints and thus extracted the
physical states is rather common in the literature. Indeed, in most
examples of constrained systems (even with just a finite
number of degrees of freedom) the inner product that
may seem ``natural'' prior to the imposition of constraints has
little physical relevance because typically the physical states have
infinite norms with respect to this inner product. However, in most
treatments, once the physical states are singled out, the inner
product is first introduced by making an appeal to a symmetry
group  --the Poincar\'e group in Minkowskian field theories-- and
physical observables are then taken to be the self-adjoint operators
on the resulting Hilbert space. In the present quantization program,
the strategy is somewhat different
\cite{book,carlolee,poona,review}.
We first solve the constraints, then introduce linear operators on
the resulting vector space that correspond to real classical
observables and in the final step seek a Hermitian inner product
that makes these  operators self-adjoint. Thus, the requirement that
real classical observables should become self-adjoint operators is
now regarded as  a {\it restriction on the choice of the inner
product}. Consequently, the program does not rely on the
Poincar\'e invariance which has  no obvious analog in
quantum gravity. Furthermore, this strategy allows us to begin with
{\it complex} canonical variables to describe the
{\it real} gravitational field. In the classical theory, one
recovers the real theory by imposing the reality conditions on the
canonical variables. In the quantum theory, these conditions dictate
Hermitian adjointness relations and therefore also the
choice of the inner product. Thus, our program is an extension of
the Dirac method of quantizing constrained systems in that it
enables us to use more general canonical variables --  such as the
real triads but complex complex connections-- and, at the same
time, contains a concrete proposal to select the required inner
product. (For further details on the general program, see
\cite{poona}.)

Let us return to linearized gravity. In this subsection, we will show
that the reality conditions do indeed suffice to pick out an  inner
product on the space ${\cal S}_{Phy}$ of physical states. Moreover,
eventhough no explicit appeal is made to Poincar\'e invariance, the
inner product we find is the correct one: the final theory is completely
equivalent to the theory of gravitons obtained via standard methods.
The same procedure has also  been shown to work in a variety of
systems, such as the quantum Maxwell theory \cite{maxwell}, $2+1$
gravity \cite{2+1}, and certain model systems (with a finite number
of degrees of freedom) that mimic various features of general
relativity \cite{abhaytate}. We suspect that there is a general
theorem which captures the idea that, for a wide class of
constrained
systems, the physical
reality conditions essentially suffice to determine the inner
product on the space of physical states.

We begin by positing a general form for an inner product on
${\cal  S}_{Phy}$. The idea is to represent the regular loop
states $\psi(\vec\gamma )$ by the functionals $\phi(F^\pm(k))$
from which they arise, make the ansatz
\f
<\psi | \psi' > := \int
\prod_{k, \pm} dF^\pm (k)\ e^{-T[F^\pm (k)]}\ \overline{\phi(F^\pm)}
\ \phi'(F^\pm)\ , \ff
and then determine the ``measure'' ${\rm exp} - T[F^\pm (k)]$ on
${\cal F}^3_{Phy}$ by imposing the reality conditions. Recall that
$F^\pm (k)$ satisfy the condition:
\f
\overline{F^\pm (k,\vec{\alpha})} = F^\pm (-k, \vec{\alpha})\  .
\label{frc}
\ff
Hence, the inner product is (formally) Hermitian if and only if
$T[F^\pm (k)]$ is real-valued. To determine its functional form,
let us now impose the reality conditions. The property (\ref{frc})
of $F^\pm (k)$ implies that the first of the two reality conditions,
(\ref{rr1}), is automatically satisfied. The second, (\ref{rr2}),
holds if and only if
\f
<\psi | (\hat{B}^\pm (k) )^\dagger  |\chi > \equiv
\overline{<\chi |\hat{B}^\pm (k) |\psi>} =
<\psi| -\hat{B}^\pm (-k) - {k^2\over G}\hat{h}^\pm (-k) |\chi >.
\ff
Using the explicit action of the operators (\ref{hop},\ref{bop})
$\hat{h}^\pm(k)$ and $\hat{B}^\pm(k)$ on physical states, we
conclude that the necessary and sufficient condition for the reality
conditions to hold is:
\f
{\delta T \over \delta F^\pm (-k)} = \mp 2 |k| \hbar G  F^\pm(k).
\label{reco}
\ff
The functional $T$ is determined uniquely (up to an additive
constant) by (\ref{reco}):
\f
T= \hbar G \int d^3k |k| \left (|F^- (k)|^2 -  |F^+ (k)|^2 \right )
\ff
Thus, the final form of the inner product is:
\f
<\psi | \psi' > = \int \prod_{k, \pm}
dF^\pm (k)\ \  e^{\hbar G \int {d^3k}|k| \left (|F^+(k)|^2 -
|F^-(k)|^2 \right)}\ \
\overline{\phi (F^\pm)} \phi' (F^\pm)
\label{ip}
\ff
Let us analyze the space of normalizable physical states. Its
structure is most transparent when expressed in terms of
functionals
$\phi (F^\pm(k))$ on ${\cal F}^3_{Phy}$. On the negative helicity
sector, the inner product (\ref{ip}) is given by just a Gaussian
integral. All polynomials in $F^-(k)$ are normalizable. The
functionals that depend on $F^+(k)$ on the other hand will not be
normalizable unless they are exponentially damped. Thus, the
normalizable states are of the form $\phi(F^\pm (k))= P(F^\pm (k))\>
{\rm exp} -\hbar G \int d^3k |k||F^+(k)|^2$, where $P$ is a
polynomial. The Hilbert space of physical states will contain more
general states; as is usual in field theory, the Cauchy completion
considerably expands this space of polynomials. Finally, we note
that the description is  again asymmetric between the two
helicities.
The detailed discussion of the origin and meaning of this asymmetry
can be found in \cite{poona}.

\subsection{Quantum Hamiltonian and the graviton states}

The  classical Hamiltonian (\ref{hamiltonian}) in terms of the
physical degrees of freedom is
\begin{eqnarray}
H =  -\half \int d^3k & & \left [ \left ( h^+(-k) + {G \over k^2 }
 B^+(-k ) \right ) B^+(k) \right.
\nonumber
\\
& & \left. + \left ( h^-(-k) + {G \over k^2 }   B^-(-k)   \right )
B^-(k) \right ] .
\label{hamiltonian2}
\end{eqnarray}
The quantum Hamiltonian operator on the physical states is
therefore obtained just by replacing the classical fields $h^\pm
(k)$ and $B^\pm (k)$ by the corresponding operators with an
appropriate factor ordering. Under the inner product
defined in the previous subsection, we have:
\f
-\left (\hat{h}^\pm (-k) + {G \over k^2 } \hat{B}^\pm (-k)
\right)^\dagger = {G \over k^2 } \hat{B}^\pm(k).
\ff
The form of the Hamiltonian and these Hermitian adjoint
relations suggest that appropriate multiples of the
operators $\hat{B}^\pm(k)$ should be identified with the creation
{\it
or} the annihilation operators of gravitons. The precise multiple
is determined by the commutation relations and dimensional
considerations. The correct identifications turn out to be:

\begin{eqnarray}
{a}_+ (k) & = & - \ \sqrt{|k|\over 2\hbar G}\ \ [\hat{h}^+(-k) +
{G \over |k|^2 } \hat{B}^+(-k)], \label{c0}
 \\
(a_+(k))^\dagger &= & \sqrt{G \over 2\hbar |k|^3 }\ \ \hat{B}^+(k) ,
\label{c1}
\\
(a_-(k))^\dagger   &= & - \ \sqrt{|k|\over 2\hbar G}\ \
[2\hat{h}^-(-k) + {G \over |k|^2} \hat{B}^-(-k)]
\label{c2}
\\
a_-(k) & = & \sqrt{G \over 2\hbar |k|^3} \ \  \hat{B}^\pm(k) .
\label{c3}
\end{eqnarray}
With this assignment of the roles to the operators, the creators and
the annihilators have the correct physical dimensions,
$({\rm length})^{-\textstyle{3\over 2}}$,  and satisfy the correct
commutation relations; the only  non-vanishing
bracket is
 \f
[a_\pm (p) , (a_{\pm}(k))^\dagger  ] = \delta^3(k-p).
\ff
Furthermore, the (normal ordered) Hamiltonian operator can now
be expressed as:
\f
H= \int d^3k\  \hbar |k| \ \left [  (a_+(k))^\dagger
a_{+}(k)  +  (a_-(k))^\dagger a_{-}(k)  \right ].
\ff
The asymmetry in the identification of creators and
annihilators for the two helicities arises because there is a
difference in sign in the commutation relations between
$\hat{h}^\pm (k)$ and $\hat{B}^\pm(k)$ which in turn arises
because of our preference for self dual connections over the
anti-self dual ones.

Let us examine the action of these operators on physical states. The
vacuum $\psi_o (\vec\gamma )$ is annihilated by the operators
$a_\pm(k)$. Using the expressions (\ref{hop},\ref{bop}) of the
operators $\hat{h}^\pm(k)$ and $\hat{B}^\pm (k)$, we have:
 \f
a_-(k) \psi_o (\vec{\gamma}) = \sqrt{1\over 2\hbar G|k|^3}\ \
{\delta \over \delta F^-(-k)} \phi_o(F^\pm ) =0 ,
\ff
and
\f
a_+(k) \psi_o (\vec{\gamma}) = -\left (\sqrt{ |k|\over 2\hbar G}\ \
[|k|^{-1}{\delta \over \delta F^+(k)}
  + 2\hbar G F^+(-k) ] \right ) \psi_0 (F^\pm ) = 0.
\ff
The solutions to these equations is:
\f
\psi_o (\vec\gamma) = e^{-\hbar G \int d^3k |k|
|F^+ [k,\vec\gamma]|^2 }\ .
\label{vacuum}
\ff
Note that the exponential damping with respect to the positive
helicity states is precisely of the form needed by the form of the
inner product (\ref{ip}). The expression of the vacuum state in this
representation is thus chirally asymmetric. Again, this feature
arises because of our asymmetric choice of basic variables:
physically, the holonomy operators are tied with self dual
connections. Had we worked with positive frequency connections
rather
than self dual, the vacuum  state would have been the unit
functional for {\it both} helicities. The situation is completely
analogous in the self-dual loop representation of the Maxwell theory
as well and is not intrinsic to gravity.

The graviton states can now be constructed by acting with
the creation operators (\ref{c1}) and (\ref{c2}) on this
vacuum.  For example, the one graviton states are:
\begin{eqnarray}
\psi_{(p,+)}(\vec{\gamma}) &= (a_+(p))^\dagger \psi_o
(\vec{\gamma})
\nonumber
&=  -\sqrt{2\hbar G |p|}\ \ F^+(p)\   e^{-\hbar G \int
d^3k |F^+(k)|^2 }
\end{eqnarray}
and
\begin{eqnarray}
\psi_{(p,-)}(\vec\gamma )&= (a_- (p))^\dagger
\psi_o (\vec{\gamma})
\nonumber
=  \sqrt{2 \hbar G |p|}\ \
F^- (p)\ ^{-\hbar G \int d^3k |k| |F^+(k)|^2}.
\end{eqnarray}
By superposing these pure momentum graviton states, one can
construct
wave packets. In terms of loops, a general one-graviton state
$f_{aj}(p)
= f^+(p)m_a(p) m_j(p) + f^-\bar{m}_a(p)\bar{m}_j$ in the Fock space
can now be expressed as:
\f
\psi[\vec{\gamma}] = \left(\sum_j \oint_{\alpha^j} f_{ja}
\dot\alpha_j^a ds\ \  \right)
\psi_o[\gamma^i]
\ff
where $f_{ja}(x)$ is the Fourier transform of $f^+(p)m_j(p)m_a(p) -
f^- (p)\bar m_j(p) \bar m_a(p) $. More generally, the helicity plus,
m-graviton states are expressed by n-nomials in $F^+$ of order
m, while the helicity minus, m-graviton states are expressed by
Hermite polynomials in $F^-$ of order n (times the vacuum). Thus,
these states have the correct functional dependence on the form
factors to be formally  normalizable.

\subsection{Divergences in the ground state wave functional}

We have presented a construction of the Fock space of free
gravitons, together with the Hamiltonian (\ref{hamiltonian2}),
inner product (\ref{ip}), an algebra of creation and
annihilation operators (\ref{c0},\ref{c1},\ref{c2},\ref{c3}), and
explicit expressions for the ground state (\ref{vacuum}) and
graviton states. However, certain steps in this construction
are only formal. As we are about to show, the momentum integral in
the expression of the ground state $\psi_o(\vec\gamma )$ diverges.
Therefore, regarded as a functional of loops, the ground state --and
hence also any n-graviton state-- wave functional vanishes on all
but the trivial loop!

The origin of this problem is the following.  From a physical point of
view, the problem is simply that the form factor of any loop, seen as
a one-particle wave-function, corresponds to a non-normalizable
(improper) state.   More precisely, consider the sub-space ${\cal
F}^3_o$ of ${\cal F}^3_{Phy}$ spanned by  functions $F^\pm (k)$ on
the momentum space for which \hfil\break
$\int d^3k |k| |F^{\pm} (k)|^2$ is finite. The vacuum state
$\phi_o(F^\pm(k))$ (as well as any $n$-graviton state) is a
non-zero, well-defined functional on this space. Why is it then
that $\psi_o (\vec\gamma )$ vanishes? The reason is that the form
factors $F^{ai}(k,\vec\gamma )$ are such that $\int d^3k |k|
|F^\pm (k, \vec \gamma )|^2$ diverges for all (non-trivial) loops
$\vec\gamma$. Therefore, the image within ${\cal F}^3_{Phy}$ of the
loop space ${\cal HL}^3$ has  zero intersection with ${\cal F}_o^3$.
This is why the n-graviton states which, strictly speaking,
have support only on the subspace ${\cal F}^3_o$ of ${\cal F}^3$ get
pulled back to the  zero functional
on the loop space. Thus, the problem has to do with the
precise choice of the function space ${\cal F}$ which we left
open. To fix this choice in an appropriate fashion, we will need
to thicken out the loops and use the corresponding ``smeared out''
form factors to map the loop space ${\cal HL}^3$ to ${\cal
F}^3$. That is, to resolve the problem, we have to sharpen the
regularity conditions on loop states. We will introduce the required
smearing in section 3.7. Once this is done, we will be able to
represent the n-graviton states as functionals on smeared out loops
in an unambiguous way. Furthermore, the overall structure of the
theory after smearing will remain essentially the same as it has
been so far.

In this sub-section, however, we will continue to work with
the unsmeared loops, examine more closely the ground state wave
functional obtained above and explore the nature of the divergent
integrals involved.

Using the definition of the $F^\pm (k, \alpha )$ in terms of the
$m(k)$ and $\bar{m}(k)$ the ground state wave-function
(\ref{vacuum}) may be written
\f
\psi_0 [\vec{\gamma} ] = e^{-l_{{l_P}}^2 \sum_{i,j} \oint ds \oint
dt (\dot{\gamma_i}^a (s)) (\dot{\gamma_j}^b (t)) \ \
G_{abij}(\gamma_i(s)-\gamma_j (t)) } ,
\ff
where $l_P= \hbar G$ is the Planck length
and
\f
G_{abij}(x)
 \equiv \int {d^3 k \over (2 \pi )^3}
 |k| e^{\imath k \cdot x}
\left[\delta_{ij} - {k_i k_j \over k^2   }   - \imath \epsilon_{ijl} {k^l
\over |k|}   \right]
\left[\delta_{ab} - \imath \epsilon_{abc} { k^c \over |k|}  \right]  .
\ff
We have used the identity
\f
m_i (k) \bar{m}_j (k) = {1 \over 2} \left [
\delta_{ij} - {k_i k_j \over k^2   }   - \imath \epsilon_{ijl} {k^l
\over |k|}  \right  ],
\ff
and the fact that the $F^a[k, \gamma ]$ are transverse.  It is
straightforward to do the momentum integrals, after which we find
\begin{eqnarray}
   G_{abij}(x) =
   & - &
 {\delta_{ij} \delta_{ab} \over |x|^4 }  +
2{\delta_{ab} x^i x^j  \over |x|^6}
 \\ \nonumber
& - & {1\over 4} \left [   \delta_{ab} \epsilon_{ijk} + \delta_{ij}
\epsilon_{abc}   \right ] {\partial \over \partial x^k}
\delta^3 (x)  \\ \nonumber
&+& {3 \over 32 \pi^3 } \epsilon_{abc}
\left [ 5 {x^i x^j x^c   \over |x|^7 } -
{ \delta_{ij} x^c + \delta_{cj} x^i + \delta_{ci} x^j
\over |x|^5   }   \right ]   \\   \nonumber
 &+& {\epsilon_{ijk} \epsilon_{abc} \over 2 |x|^4    }
\left [  \delta_{ck} -  {4  x^c  x^k \over
|x|^2}  \right ]   .
\end{eqnarray}
 Now, it is easy to see that the two integrals in the definition
of the vacuum state are divergent for every non-trivial loop. Note
that the divergences occur whenever $|\gamma_i (s)-\gamma_j
(t)|=0$,
which means either at an intersection or in the diagonal terms in
the sum $\sum_{ij}$.

To clarify the nature of these divergences, let us consider the
value of the ground state wave-function on a particular loop. We
evaluate the integrals for non-intersecting loops, so that the
divergences are only in the diagonal terms. We note that many of the
possibly divergent terms can be set equal to  zero if we choose the
loops to be {\it transverse}, by which we mean  that the $i'th$ loop
always runs in a plane perpendicular to the $i'th$ direction.   We
then have, up to finite terms,
\f
\psi_0[\vec\gamma] = e^{ \sum_i W[\gamma_i] }
\ff
\f
W[\gamma] = {l_P}^2 \oint ds \oint dt {\dot{\gamma}^a (s)
\dot{\gamma}^b (t)
\delta_{ab} \over |\gamma(s)-\gamma(t)|^4 }
\ff
We expand $\gamma^a  (t)=\gamma^a (s) + (t-s) \dot{\gamma}^a (s)
+...$ and, for convenience,  choose a parametrization such that
$|\dot{\gamma}^a|=1$ (so that $\dot{\gamma} \cdot
\ddot{\gamma}=0$).
We find
\f
W[\gamma] = - l_{{l_P}}^2 \left [ \oint_0^L  ds \oint_0^L  dt
 {1 \over (t-s)^4 } - \oint_0^L  ds  |\ddot{\gamma}(s) \cdot
\ddot{\gamma}(s)|
\oint_0^L  dt {1 \over 2 (t-s)^2 } + ...   \right ]
\ff
We can draw two conclusions from this expression.  First, the
leading divergence is independent of the loop.  Were this the only
divergence (as is the case in Maxwell theory \cite{maxwell})
then the divergence could be removed by a wave-function
renormalization of the form,
\f
\psi_0^{Ren}\equiv e^{ l_{{l_P}}^2 \oint_0^L  ds \oint_0^L
dt
 {1 \over (t-s)^4 } }\ \psi_0
\ff
(defined in the context of an appropriate regularization procedure.)
However, there are non-leading divergences which depend on the
loop, and,
in particular, on the averages of higher derivatives of the loop
taken around the loop. Indeed, we see that even if wave-function
renormalization is done, the renormalized ground state will still
vanish on all loops except the straight lines.

The difference between the case of Maxwell theory and linearized
gravity may be traced directly to the dimensional nature of the
gravitational constant.  Because of the presence of $\hbar G$, the
momentum integrals are divergent like $l_{{l_P}}^2 /\epsilon^2$,
where $\epsilon$ is any short distance cutoff.  In the Maxwell case,
where there is no dimensional constant, the corresponding
expression
is only logarithmically divergent, so that the only divergences are
loop independent.

\subsection{Averaging}

In this section, we construct a modified version of the
representation defined above in which  no divergent quantity
appears. (Further details can be found in \cite{maxwell}.)

To motivate the modification needed, let us first make a brief
detour to note that we do already have a well-defined
representation of the algebra of the the graviton operators
$\hat{h}^\pm (k)$ and the magnetic field operators
$\hat{B}^\pm(k)$ in which the states are represented by
functionals $\phi (F^\pm(k))$ on the space ${\cal F}^3_o$. (Recall
that this is the  space spanned by $F^\pm(k)$ for which $\int d^3k
|k| |F^\pm(k)|^2$ is finite.) On this space, the inner product is
given by the  right side of (\ref{ip}),
\f
<\phi|\phi'> = \int \prod_{k,\pm} dF^{\pm}(k)\ e^{\hbar G
\int d^3k |k|(|F^+(k)|^2 -|F^-(k)|^2)}\ \ \overline{\phi (F^\pm)}
\phi' (F^\pm)\ ,
\ff
The vacuum state by the right side of (\ref{vacuum}),
\f
\phi_o(F^\pm ) = e^{-\hbar G \int d^3k |k| |F^+(k)|^2} \ ,
\ff
the operators $\hat{h}^\pm(k)$ act via the right side of (\ref{hop})
and
\f
\hat{h}^\pm(k)\cdot \phi (F^\pm ) = 2 \hbar G F^{\pm}(k)\ \phi(F^\pm)
\ff
while the operators $\hat{B}^\pm(k)$ act via the right side of
(\ref{bop})
\f
\hat{B}^\pm (k)\cdot \phi (F^\pm ) = \pm {|k|\over G}\ {\delta\over
\delta F^\pm(-k)}\cdot \phi (F^\pm ).
\ff
{\it The problem is that of transferring this information
to the loop space.} The strategy we adopted in section 3.2 was to use
the form factors $F^{a}(k,\gamma^i)$ to map the loop space ${\cal
HL}^3$
into the space ${\cal F}^3$ spanned by triplets of vector fields
$F^{ai}(k)$ in the momentum space, and ``pull back'' the functionals
$\phi(F^{ai}(k))$ on  ${\cal F}^3$ to functionals $\psi (\vec\gamma
)$ on the loop space via $\psi (\vec\gamma ) = \phi (F^{ai}(k) =
F^a(k, \gamma^i ))$. As remarked earlier, however, under this map
the image of ${\cal HL}^3$ in
${\cal F}^3$ has no intersection with the space ${\cal F}^3_o$
since the integrals $\int d^3k |k||F^\pm (k, \vec\gamma)|^2$
diverge for all non-trivial loops $\vec\gamma$.
Therefore, the pull-back of the n-graviton states $\phi (F^\pm (k))$
to the loop space vanishes identically. The idea now is to modify the
mapping from the loop space ${\cal HL}^3$ into ${\cal F}^3$ by
thickening the loops in an appropriate fashion.

Let us fix an universal averaging function $f_r(y)$, such that its
integral over the three-dimensional space is 1 and that it goes to a
delta function when $r\rightarrow 0$. For concreteness, we make
the
following choice: \f
     f_r(y) = {1\over (2\pi)^{3/2} } e^{-{y^2\over 2r^2} }.
\ff
Given any loop $\gamma$ we consider the loop $\gamma+ y$
obtained by
rigidly shifting the loop by a distance $y$:
\f
(\gamma+y)^a(s) = \gamma^a(s)+y^a,
\ff
and we average over $y$ using the weight $f_r(y)$. In particular,
we define the ``average'' of the distribution with support on
the loop defined in equation (\ref{loopdistr}) as
\begin{eqnarray}
   F^a_r [x,\gamma ] &=& \int d^3y f_r(y) F^a[x,\gamma+ y]
\nonumber \\
 & = & \int d^3y \ f_r( y) \ \oint ds \ \dot\gamma^a(s)\ \delta^3( x-
\gamma(s) +y) \nonumber \\
 & = &  \oint ds \ \dot\gamma^a(s)\ f_r(x-\gamma(s)).
\end{eqnarray}
Its Fourier transform $F^a_r[k, \gamma]$ is by definition the
{\it averaged form factor}. It is easy to verify that this turns out
to be the product of the standard form factor and the Fourier
transform of $f_r$: \f
    F^a_r[k, \gamma] =  e^{-{r^2k^2\over 2}} F^a[k, \gamma];
\ff
the averaging on the small tube in coordinate space corresponds to a
damping factor for high momenta. We note that since for any
(finite, smooth) loop $\gamma$, the averaged form factor $F^a_r[k,
\gamma]$ is the Fourier transform of a smooth vector field with
compact support, the integral $\int d^3k |k||F^a_r [k, \gamma]|^2$
converges. This technical difference between the averaged form
factors and the ``bare'' form factors introduced earlier plays
the key role in what follows.

We now use these averaged form factors $F^a_r [k, \gamma]$ to
define
a map from ${\cal HL}^3$ into ${\cal F}^3$ in the obvious fashion:
$ [\vec\gamma ] \mapsto F^{ai}(k) := F^a_r[k, \gamma^i]$. Denote by
${\cal H}_F$ the (pre-)Hilbert space of all functionals $\phi
(F^\pm(k))$ on ${\cal F}_o$ which have finite norm under (\ref{ip}).
These functionals can now be pulled back to ${\cal HL}^3$. Denote
by ${\cal H}_L$ the space of functionals $\psi_r (\vec\gamma)$
so obtained on the loop space; $\psi_r(\vec\gamma) =
\phi(F^\pm_r(k))$
for some $\phi \in {\cal H}_F$. This is the required space of
physical loop states. Thus, now, in the loop representation, the
vacuum is given by:
\f
(\psi_{r})_o(\vec\gamma) = e^{-\hbar G \int d^3k |k|
|F_r^+[k,\vec\gamma ]|^2 }.
\label{ipa}
\ff
Now, the integral in the exponent is manifestly finite and the
vacuum wave functional is well-defined on the entire loop space. In
the terminology used in section 3.6, the averaging procedure has
replaced the divergent expression (\ref{vacuum}) for the vacuum
loop
functional by the well defined loop functional
\f
\psi_{r\,0} [\vec{\gamma} ] = e^{-l_{{l_P}}^2 \sum_{i,j} \oint ds
\oint dt
\dot{\gamma}^a_i (s) \dot{\gamma}^b_j (t) G_{abij}^{(r)} (\gamma_i
(s) -
\gamma_j (t) )},         \label{vacuum2}
\ff
where
\f
G_{abij}^r (x) = \int {d^3 k \over (2 \pi )^3 }\ e^{\imath k \cdot x}\
G_{abij}^r (k),
\ff \f
\tilde{G}_{abij}^r (k) = e^{-r^2k^2}   |k| \left[\delta_{ij} - {k_i k_j
\over
k^2 }   - \imath \epsilon_{ijl} {k^l  \over |k|} \right]
\left[\delta_{ab} -
\imath \epsilon_{abc} { k^c \over |k|}  \right] .
\ff
Indeed, this is the only change which is brought about by the
averaging procedure. The wave functionals of 1-graviton states are
now essentially the same as before; the only difference is that
the form factors are replaced by their average values. The inner
products between these states are all manifestly finite since all
the momentum integrals involved converge. Next, let us
consider the physical operators $\hat{h}^\pm(k)$ and
$\hat{B}^\pm(k)$. Their expressions are again essentially the same;
the only change is that in the action (\ref{hop}) of the graviton
operators, the form factor is replaced by its averaged value
and in the action (\ref{bop}) of the magnetic field operators, the
functional derivative is evaluated at the averaged form factor.

Thus, for each finite, non- zero value of $r$, we have a well-defined
representation of the quantum field theory in which states arise
as functionals on the loop space. There are no infinities and,
irrespective of the choice of the averaging parameter $r$, the
representation is entirely equivalent to the standard Fock
representation. There is no cut-off involved and the loop
description captures gravitons with arbitrarily high frequencies. In
any calculation of a physical quantity, the dependence on $r$
cancels out. Thus, $r$ is not a regularization parameter.

\section{Discussion}

We have completed the main goal of this paper, which is the
construction of the Fock space of graviton states, together
with inner product and physical observables, in terms of the
loop quantization.  The vacuum is given by the loop functional
(\ref{vacuum2}), the n-graviton states are given by the loop
functionals obtained by multiplying the vacuum with powers
of  $F_r^+[k,\vec\alpha]$ and Hermite polynomials
in $F_r^-[k,\vec\alpha]$. The inner product is given by the
expression (\ref{ipa}). For the physical states this expression
reduces to Gaussian integrals of polynomials. The Hilbert space of
states $\cal H$ is formed by the loop functionals
$\psi_r[\vec\alpha]$  which are normalizable linear combinations of
the n-graviton states. The structure that we have obtained is
isomorphic to the standard Fock space of gravitons.

\noindent
We conclude with a few remarks.

- In the construction of the theory we have used a somewhat novel
quantization procedure: the connections $A_{aj}$ are self-dual
rather than of positive frequency and the inner product on the
physical states is selected using the reality condition. The
inner product so determined turns out to be the standard Poincar\'e
invariant one in spite of the fact that Poincar\'e invariance is
never explicitly invoked. These features are important from the
viewpoint of full quantum general relativity where we cannot define
positive frequency fields nor do we have an access to the Poincar\'e
group.

- Two peculiar features of the loop representation of linearized
gravity are the surprising chiral asymmetry of the formalism and
the use of {\it triplets} of multiple loops. The first is a
consequence of the chiral asymmetry of the new canonical
framework;
the second is the key technical idea that allowed us to construct
the theory.

-Note that the Planck length $l_P$ remains a parameter in
the quantum theory of free gravitational waves. This is not
surprising. The situation is analogous to that in particle
mechanics where, although the mass does not appear in the classical
equations of motion of a free particle, the Compton wave-length {\it
is} an important parameter in the corresponding  quantum theory.

-The averaged representation depends on an arbitrary parameter
$r$. (More precisely, it depends upon an arbitrary function
$f_r(x)$.)  Physical quantities do not depend on this
parameter. The physical meaning, if any, of this parameter
remains unclear. Had we worked with positive frequency connections
rather than self dual, for example, the averaging would not have been
necessary. Recall also that, unlike in the Maxwell theory, wave
function renormalization does not remove divergences in the present
description of linearized gravity. This difference is, in turn, due
to the fact that the linearized theory of gravitons still contains a
dimensional  parameter,  the $l_P$ length.  As a result, the
momentum integrals in the ground state  wavefunctionals
(and inner product) are now two powers of momentum more
divergent than they are in the Maxwell theory.
These facts are, of course, related to the non-renormalizability
of the standard  perturbation theory for general
relativity. In the loop representation geared to self-dual
connections, they seem to have consequences already at the linear
level.

- The loop representation is equivalent to the Fock representation
for linearized gravity. Hence, if we attempt to construct the
conventional perturbation theory using the framework of this
paper, we would find the same results as in any other treatment
of the Fock states. The potential interest of our
results is in another direction. From this standpoint, the relevant
question is the following: Is the Fock space of
gravitons the correct linearization of the full quantum theory of
general relativity? In the full non-perturbative quantization of
general relativity  there should  be states that represent classical
field configurations, at least at large scales. This follows from
the correspondence principle.  However, there is no principle that
guarantees that these exact quantum states that look
semi-classical at large scales continue to  behave semiclassically
at the  Plank scale. If the full quantum theory implies
that at short distances there is no semiclassical
behavior, then we should expect that the quantization of the
linear theory around a flat background space does not make sense at
short distances. Since in the loop representation we have
solutions of the full non-perturbative quantum constraints
\cite{carlolee} and since we now understand the linearized theory in
the loop representation, it should be possible to
investigate this issue in detail. Work is now in progress along
these lines
\cite{linearization}.

\bigskip

A.A. and L.S. thank the Department of Physics of the University of
Trento, and in particular Marco Toller, for the hospitality
during the period in which this work was completed. This research
was
supported in part by the NSF grants INT88 15209, PHY90 16733 and
PHY90
12099 and by research funds provided by Syracuse University.


\begin{thebibliography}{33}

\bibitem{abhay} A.~Ashtekar, Phys. Rev. Lett. 57,
2244 (1986); Phys. Rev. D36, 1587 (1987).

\bibitem{tedlee}  T.~Jacobson and L.~Smolin, Nucl. Phys. B299, 295
(1988).

\bibitem{carlolee} C.~Rovelli, L.~Smolin, Nucl. Phys. B133 (1990)
80; Phys. Rev. Lett. 61, 1155 (1988).

\bibitem{poona} A.~Ashtekar , Lectures on {\it Non-perturbative
canonical gravity}, notes prepared in collaboration with
R.S.~Tate (World Scientific, Singapore, 1991).

\bibitem{review} C.~Rovelli,  Ashtekar formulation of
general relativity and loop-representation of quantum
gravity: a report, Class. and Quant. Grav. (in press).

\bibitem{gambini} R.~Gambini and A.~Trias, Phys. Rev. D23 553
(1981); Lett. al Nuovo Cimento 38, 497 (1983); Phys. Rev. Lett. 53,
2359 (1984); Nucl. Phys. B278, 436 (1986); Phys. Rev. D39, 3127
(1989); R.~Gambini, Loop space representation of quantum general
relativity and the group of loops, Preprint University of
Montevideo, (1990).

\bibitem{2+1} A.~Ashtekar, V.~Husain, C.~Rovelli, J.~Samuel and
L.~Smolin , Class. and Quan. Grav. L185-L193 (1989).

\bibitem{abhaytate} A.~Ashtekar and R.S.~Tate, An extension of the
Dirac program for quantization of constrained systems: Examples,
in preparation (1991).

\bibitem{maxwell}
A.~Ashtekar and C.~Rovelli, A loop representation for the quantum
Maxwell field, Syracuse preprint
(1991).

\bibitem{yangmills}   C.~Rovelli and L.~Smolin, Loop
representation for  lattice gauge theory, Pittsburgh and
Syracuse preprint (1990).

\bibitem{aa}  A.~Ashtekar, J. Math. Phys. 27, 824 (1986);
A.~Ashtekar, C.~Rovelli and L.~Smolin,
Self duality and quantization, J. Geo. \& Phys. (in press).

\bibitem{penrose} R.~Penrose and W.~Rindler, {\it Spinors and
spacetime}, Vol. 2 (Cambridge University Press, Cambridge 1986).

\bibitem{tedcarlos}  C.~Kozameh, W.~Lamberti and T.~Newman,
 Holonomies and Einstein Equations, (to appear in  Ann. of Phys.
1991).

\bibitem{aajoohan} A~Ashtekar and J.~Lee,  Weak field limit of
general relativity: A new Hamiltonian description. Syracuse
preprint
(1991).

\bibitem{selfdual} A.~Ashtekar, L.~Smolin and C.~Rovelli, Self duality
and quantization of linearized gravity, in  preparation
(1991).

\bibitem{googly}   L.~Smolin,  The $G_{Newton} \rightarrow 0$
limit  of Euclidean quantum gravity, in preparation.

\bibitem{book} A.~Ashtekar (with invited contributions), {\it New
Perspectives in Canonical Gravity}, (Bibliopolis, Napoli, Italy,
1988).

\bibitem{time} C.~Rovelli,  Time in quantum gravity: Physics
beyond the Schr\"odinger regime, Universita' di Roma preprint
(1988); Phys. Rev. D43, 442 (1991).

\bibitem{linearization} A.~Ashtekar, C.~Rovelli and L.~Smolin, in
preparation (1991).


\end{thebibliography}
\end{document}